# Machine Learning and Glioma Imaging Biomarkers

**Authors:** Thomas C Booth[1,2,✉], Matthew Williams[3], Aysha Luis[1,4], Jorge Cardoso[1], Keyoumars Ashkan[5], Haris Shuaib[6,7]

**Affiliations:**

[1]King's College London, School of Biomedical Engineering & Imaging Sciences, St. Thomas' Hospital, London. SE1 7EH. United Kingdom.

[2]Department of Neuroradiology, King's College Hospital NHS Foundation Trust, London. SE5 9RS. United Kingdom.

[3]Department of Neuro-oncology, Imperial College Healthcare NHS Trust, Fulham Palace Rd, London, W6 8RF. United Kingdom.

[4]Department of Radiology, *St George's* University Hospitals *NHS Foundation* Trust - Blackshaw Road, SW17 0QT London, United Kingdom.

[5]Department of Neurosurgery, King's College Hospital NHS Foundation Trust, London. SE5 9RS. United Kingdom.

[6]Department of Medical Physics, Guy's & St. Thomas' NHS Foundation Trust, London. SE1 7EH. United Kingdom.

[7]Institute of Psychiatry, Psychology & Neuroscience, King's College London, London, SE5 8AF. United Kingdom.

**Declaration of Interest:** Jorge Cardoso is involved in machine learning enterprise and business

**Acknowledgments** This work was supported by the Wellcome/EPSRC Centre for Medical Engineering [WT 203148/Z/16/Z].

✉**Corresponding author:** tombooth@doctors.org.uk
King's College London, School of Biomedical Engineering & Imaging Sciences, St. Thomas' Hospital, London. SE1 7EH. United Kingdom.

# Machine Learning and Glioma Imaging Biomarkers

Background: Increased computational processing power and advances in database curation will facilitate the development of biomarkers that may contribute to the defeat of cancer in the mid-21$^{st}$-century.

Aim: We review how machine learning is applied to imaging biomarkers in neuro-oncology, in particular for diagnosis, prognosis and treatment response monitoring.

Methods: The PubMed and MEDLINE databases were searched for articles published before September 2018 using relevant search terms. The search strategy focused on articles applying ML to high grade glioma biomarkers for treatment response monitoring, prognosis and prediction.

Results: Magnetic resonance imaging is typically used throughout the patient pathway because routine structural imaging provides detailed anatomical and pathological information and advanced techniques provide additional physiological detail. Using carefully chosen image features, machine learning is frequently used to allow accurate classification in a variety of scenarios. Rather than being chosen by human selection, machine learning also enables image features to be identified by an algorithm. Much research is applied to determining molecular profiles, histological tumour grade and prognosis using magnetic resonance images acquired at the time that patients first present with a brain tumour. Differentiating a treatment response from a post-treatment related effect using imaging is clinically important and also an area of active study (described here in one of two Special Issue publications dedicated to the application of machine learning in glioma imaging).

Conclusion: Whilst pioneering, most of the evidence is of a low level having been obtained retrospectively and in single centres. Studies applying machine learning to build neuro-oncology monitoring biomarker models have yet to show overall advantage over those using traditional statistical methods. Development and validation of machine learning models applied to neuro-oncology require large, well-annotated datasets, and therefore multidisciplinary and multi-centre collaborations are necessary.

# Machine Learning and Glioma Imaging Biomarkers

**Introduction**

A biomarker, a portmanteau of biological and marker, is defined as a characteristic that is measured as an indicator of normal biological processes, pathogenic processes, or responses to a therapeutic intervention[1]. Molecular, histologic, imaging, or physiologic characteristics are types of biomarkers. Well-known biomarkers in neuro-oncology include demographic features (such as age) and tumour features (such as grade and $O^6$-methylguanine-DNA methyltransferase (MGMT) promoter methylation status), while imaging biomarkers are used for diagnosis, prognosis and treatment response monitoring.

MRI is typically used throughout the neuro-oncology patient pathway because routine structural imaging provides detailed anatomical and pathological information, and advanced techniques (such as 1H-magnetic resonance spectroscopy) provide additional physiological detail[2]. Qualitative analysis of a new intracranial mass aids diagnosis and in routine clinical practice can determine whether or not to proceed to confirmatory biopsy or resection. For example, with some basic demographic information such as the age of the patient and with some clinical information, such as knowledge that the mass was found incidentally whilst investigating an unrelated condition, the qualitative routine structural imaging features of a grade 1 meningioma allow diagnosis with high precision (positive predictive value) without the need for confirmatory biopsy. Advanced imaging techniques allow quantitative analysis of abnormalities that can change management. For example, cerebral blood volume values obtained using dynamic susceptibility-weighted contrast-enhanced imaging (DSC) imaging within an area of tumour contrast enhancement, or 1H-magnetic resonance spectroscopic ratios acquired from a mass, may help determine whether a tumour is of high histological grade (grade 3 or 4) in certain scenarios.

Some image analysis recommendations, which determine treatment response of high histological grade gliomas (Box 1), have become common in the research setting and rely on simple linear metrics of image features, namely the product of the maximal perpendicular cross-sectional dimensions of contrast enhancing tumour (in "measurable" lesions which are defined as beyond 10 mm in all perpendicular dimensions)[3,4]. Nonetheless, seemingly simple measurements can still be challenging because tumours have a variety of shapes, may be confined to a cavity rim, and the edge may be difficult to define. Indeed, large, cyst-like high grade gliomas are common and are often "non-measurable" unless a solid peripheral nodular component fulfils the above "measurable" criteria.

Box 1. Neuro-oncology epidemiology

> The global incidence of CNS tumours is unknown but is at least 45/100 000 patients a year[5,6]. CNS tumours are categorised as primary or secondary. Secondary CNS tumours (metastases) are the commonest type of CNS tumour in adults. The reported incidence of metastatic CNS tumours is increasing but the exact incidence is unknown. Primary CNS tumours are diverse histological entities with different causes and include malignant, benign and borderline tumours. The 2016 World Health Organization classification of primary CNS tumours is based on histopathological and molecular criteria[7]. In the USA, the incidence of primary CNS tumours is 21/100,000 patients a year[8]. The two main histological types are meningiomas and gliomas accounting for 36% and 28% of primary CNS tumours respectively.
>
> There are 4 histological glioma grades. Grade 4 gliomas (glioblastoma) are the commonest glioma (53%)[9]. Diffuse grade 2 (diffuse low-grade) and 3 (anaplastic) gliomas account for approximately 30% of all gliomas. The median age at diagnosis of these gliomas are 64, 43 and 56 years respectively. In contrast, the commonest paediatric gliomas are grade 1 (predominantly pilocytic astrocytomas) accounting for 33% of paediatric gliomas[10]. Almost all machine learning studies applied to neuro-oncology have focused on gliomas, particularly high grade gliomas (grades 3 and 4) which are the malignant gliomas.

Much research in image analysis aims to extract underlying quantitative information from the imaging dataset to develop biomarkers that may not be readily visible to individual human raters; this is *radiomics*. Typically, radiomics consists of the following phases: pre-processing images, feature estimation (quantifying or characterizing the image), feature selection (dimensionality reduction to remove noise and random error in the underlying data and therefore reduce overfitting), classification (decision or discriminant analysis) and evaluation[11] (Fig. 1). Pre-processing typically constitutes a major part of most studies. Although many steps can be taken prior to patient imaging to reduce the pre-processing burden (e.g. overcoming geometric distortion through phantom analysis or reduce image noise through signal averaging), typically images will require intensity non-uniformity correction (through estimation of bias field), noise reduction (through careful application of filters), motion correction, and intensity normalization (through transformation of intensity to standard scale), and often spatial normalization (different brains anatomically aligned through geometrical transformation), and segmentation. Pre-processing pipelines are complex but potentially can have empirical, data-driven, and complete machine learning solutions to the problems described above[13], including quantification of the inherent uncertainty[14].

Some research has leveraged applied statistical models, some machine learning (ML) models and many both. The basic difference between them is that statistics draws population inferences from a sample, and ML finds generalizable predictive patterns[15]. Some of the recent shifts towards ML can be attributable, firstly, to ML methods being effective when applied to 'wide data', where the number of input variables exceeds the number of subjects; and secondly, to applied statistical modeling being inherently designed for data with tens of input variables and sample sizes smaller than those seen with current data curation (big data). Together, these explain some of the recent shifts towards ML. In this review we focus on ML approaches to neuro-oncology radiomics (Box 2).

Box 2. Assessing machine learning methodology in neuro-oncology radiomic studies

> One of the challenges when interpreting the literature on machine learning (ML) approaches to neuro-oncology is that different researchers may use different technologies as the basis for their work. As a result, the reader can face technical details that may appear challenging. In fact, many techniques share similar underlying motivations, and even when they do not, there are some basic principles that apply to assessing ML applications. Firstly, because ML models tend to start with the data and then generalise, *overfitting* is a substantial challenge. For this reason, model validation on dual training and testing datasets is recommended. Secondly, common, simple clinical data incorporation or comparison is likely to be important. Thirdly, assessing performance against an existing standard (typically an existing assessment system or human expert performance) is essential.

There has been a long history of using ML in neuro-oncology, and even neural networks have been applied to classifier tasks for more than two decades[16]. However, recent work has made use of improvements in technology to allow the use of much more complex supervised, unsupervised and reinforcement ML including the use of deep (multiple layered) neural networks (some relevant open source tools are listed in Supplementary Box 1). Nonetheless, for now, most radiomic work uses explicit rather than implicit feature engineering techniques (i.e. features chosen by imaging scientists such as texture[17], rather than features identified by an algorithm).

Evaluation in image analysis research initially consists of *analytical validation*, where accuracy and reliability of the biomarker are assessed[18]. Accuracy determines how often a test is correct in a given population (the number of true positives and true negatives divided by the number of overall tests). Accuracy alone is limited and other metrics derived from the confusion matrix are typically employed such as precision (positive prediction value), recall (sensitivity), the F1 score (recall and precision combined), balanced accuracy (the mean of sensitivity and specificity) and area under the receiver operator curve (AUC). *Clinical validation* is the testing of biomarker performance, typically in a clinical trial. One weakness of much current work is that novel approaches are validated against existing biomarkers. For example, an attempt to validate a new DSC imaging biomarker for treatment response monitoring may involve benchmarking it against a common biomarker for treatment response, such as the product of the maximal perpendicular cross-sectional dimensions of contrast enhancing tumour, rather than overall survival. However, the common biomarker itself may not be rigorously proven to be clinically valid. Indeed, when the maximal perpendicular cross-sectional dimensions of contrast enhancing tumour have been used to determine progression-free survival in high grade glioma, there may be false positive progression (pseudoprogression described below) or, when bevacuzimab is added to the treatment regimen, false negative progression (pseudoresponse). Even expert recommendations[4] for avoiding false positive progression through careful timing of cross-sectional measurements are flawed, requiring modifications[12]. False negative progression is a concern in the United States but rarely in Europe as the European Medicines Agency concluded that the progression-free survival bevacuzimab trial outcome measures were inherently confounded and the use of bevacuzimab is not supported[19].

This review describes several illustrative radiomic studies aimed at developing imaging biomarkers for treatment response monitoring, prognosis and prediction as well as diagnosis (outlined in the adjoining publication: Deep learning can see the unseeable: predicting molecular markers from MRI of brain gliomas). We demonstrate how different ML strategies are used in classification in particular, as well as in feature estimation and selection. As is fundamental to biomarker development, the extent of analytical and clinical validation is highlighted. The studies described here, many of which are retrospective and performed in single centres, show that while there is considerable research on applying ML to neuro-oncology, the evidence is often poor thereby limiting clinical utility and deployment[20].

## Material and Methods

The PubMed and MEDLINE databases were searched for articles published between September 2008 and 2018 (reviews) and September 2013 and 2018 (original research) using the search terms listed in Supplementary Table 1 based on variants of glioma and machine learning search term combinations. We excluded those articles where there was no mention of a machine learning algorithm used in feature extraction, selection or classification/regression. We excluded all articles which were not in the English language or did not have an obtainable English language translation. We excluded all articles which had no mention of imaging in the abstract or title.

Given that our review describes a broad range of studies involving several imaging approaches (a range of MRI sequences including structural and advanced techniques; also PET) and several target conditions (pseudoprogression, radiation necrosis, or a combination of both; complete response) it is not suitable for a PRISMA-DTA analysis addressing a specific question on diagnostic accuracy[21]. Nonetheless, components of the PRISMA-DTA methodology have been incorporated where practicable.

## Results

The search strategy returned 1549 initial candidate articles. Following our exclusion steps (Supplementary Figure 1) the final dataset consisted of 20 studies primarily assessing prognostic biomarkers and 14 studies primarily assessing monitoring biomarkers.

## Monitoring Biomarkers

Monitoring biomarkers are measured serially and may detect change in extent of disease, provide evidence of treatment exposure or assess safety[1]. There is an overlap with safety biomarkers which specifically determine any treatment toxicity. Monitoring blood or cerebral spinal fluid for circulating tumour cells, exosomes, and microRNAs shows promise[18]. However, imaging is particularly useful as it is non-invasive and captures the entire tumour volume and adjacent tissues and has led to recommendations to determine treatment response in trials[3,4]. Clinical validation is typically not proven. Common biomarkers are frequently used as benchmarks in an attempt to indirectly validate the monitoring biomarker under development.

The commonest primary malignant brain tumour, glioblastoma, is a devastating disease with a progression free-survival of 15% at 1 year and a median overall survival of 14.6 months despite standard of care treatment[22,23]. The standard of care treatment consists of maximal debulking surgery and radiotherapy, with concomitant and adjuvant temozolomide[22], but is associated with pseudoprogression. This describes false-positive progressive disease within 6 months of chemoradiotherapy, typically determined by changes in contrast enhancement on $T_1$-weighted MR images, representing non-specific blood-brain barrier disruption[24,25] (Fig. 2). Pseudoprogression confounds response assessment and may affect clinical management. It occurs in 20-30% of cases and is associated with better clinical outcomes[26]. Apparent tumour progression on MRI, therefore, commonly presents the neuro-oncologist with the difficult decision as to whether to continue adjuvant temozolomide or not. An imaging technique that reliably differentiates patients with true progression from those with pseudoprogression would allow an early change in treatment strategy with cessation of ineffective treatment and the option of implementing second line therapies[27]. This is an area of significant potential impact: only 50% of patients with glioblastoma receive second-line treatment, even in clinical trials.

Pseudoprogression is an early-delayed treatment effect, in contrast to the late-delayed radiation effect (or radiation necrosis)[28]. Whereas pseudoprogression occurs during or within 6 months of chemoradiotherapy, radiation necrosis occurs after this period but at an incidence that is an order of magnitude smaller than the earlier pseudoprogression. In the same way that it would be beneficial to have an imaging technique that discriminates true progression from pseudoprogression, an imaging technique that discriminates true progression from radiation necrosis would also be beneficial to allow the neuro-oncologist to know whether to implement second line therapies or not.

For these reasons, multiple radiomic studies have attempted to develop monitoring biomarkers and ML has been central to the method (Table 1). We describe several of these studies below in order to demonstrate a range of ML techniques which incorporate different imaging approaches (e.g. different sequences and combinations of sequences) and serve as examples containing methodological strengths and weaknesses. Other monitoring biomarkers have been developed for other reasons including surveillance imaging of low grade gliomas which will invariably transform to a high grade glioma[29].

Going solo: a single imaging type can be used to analyse pseudoprogression

In the first example, the study aim was to use an ML algorithm to differentiate progression from pseudoprogression in glioblastoma at the earliest time point when an enlarging contrast-enhancing lesion is seen within 6 months following chemoradiation completion, using $T_2$-weighted images alone[12]. Unsupervised feature estimation was performed to investigate topological descriptors of image heterogeneity (Minkowski functionals). Confounders were determined using principal component analysis and they showed that simple clinical features (e.g. Karnofsky performance status), were not discriminatory. Feature selection reduced the number of features to consider from 32 to 7. Supervised analysis with a support vector machine (SVM) and leave-one-out cross validation (LOOCV) gave an accuracy of 0.88 and AUC of 0.9 in a retrospective training dataset of 17 patients and the model gave 0.86 accuracy in a prospective test dataset of 7 patients with 100% recall and 80% precision. Although not apparent to the reporting radiologist, the $T_2$-weighted hyperintensity phenotype of those patients with progression was heterogeneous, large and frond-like when compared to those with pseudoprogression. The pseudoprogression phenotype on $T_2$-weighted images was shown to be a distinct entity and different from vasogenic oedema and radiation necrosis.

Additional analytical validation was performed firstly in the form of reliability testing which showed that a different operator performing segmentation achieved 100% classification concordance. Secondly, the same results using a different software package and a different operator were also obtained. Thirdly, a different feature selection method (random forest) and classifier (least absolute shrinkage and selection operator; LASSO) were used and also gave the same evaluation values with 6 similar selected features.

A strength of the study is that $T_2$-weighted images alone were used increasing the chance of translation to the clinic. However, the study was small and performed in a single centre and the biomarker requires clinical validation in a larger multicentre test dataset (open access code was provided for others to study this).

In another study, the aim was also to use an ML algorithm to differentiate progression from pseudoprogression at the earliest time point when an enlarging contrast-enhancing lesion is seen, using [18F]-fluoroethyl-L-tyrosine (FET) positron emission tomography[30]. The small, single centre, proof-of-concept study which included all high grade gliomas, showed that ML could be applied to imaging modalities other than MRI. First and second order statistics were obtained from the images of 14 patients and underwent unsupervised consensus clustering. The cumulative distribution function was used to determine the optimal class size. Feature selection by predictive analysis of microarrays methodology using 10-fold cross validation reduced the features from 19 to 10. Three class PET-based clusters were derived and progression and pseudoprogression could be differentiated with 90% recall and precision. However, there was no test dataset and the performance was similar to standard PET analysis using the maximal tracer uptake in the tumour divided by that in normally appearing brain tissue. This study highlights some of the challenges with such studies: the sample size is small, and there is no clear proof that the new approach is better than existing ones.

Another glioblastoma study aimed to differentiate progression from pseudoprogression at the earliest time point when an enlarging contrast-enhancing lesion is seen, using post-contrast $T_1$-weighted images alone[31]. They constructed a convolutional neural network (CNN) using data from 59 patients and tested its performance in 19 patients. The model performed better when combined with clinical parameters than without, giving an AUC of 0.83, area under the precision-recall curve (AUPRC) of 0.87, and F1-score of 0.74. As is the case with much CNN-based work, they were unable to determine what features were important among the input data. The optimal CNN model also performed better than a random forest model with clinical parameters alone, although it is worth noting that performance status was not included[32]. The strengths were that the testing dataset came from a second hospital and that it used post-contrast $T_1$-weighted images alone, which makes the approach potentially more applicable. Again, open access code is provided.

In summary, the three studies above demonstrate that a range of ML techniques can be used to differentiate progression and pseudoprogression using a single imaging type alone (whether $T_2$-weighted or post-contrast $T_1$-weighted MRI images or FET images) thereby increasing the chance of translation to the clinic. The importance of carefully crafting the clinical methodology in ML applications is highlighted in the CNN and FET studies described above, because the aim to differentiate progression and pseudoprogression was not truly addressed. This is because pseudoprogression and radiation necrosis (late-delayed radiation effects) are not interchangeable terms[28]. Although some researchers have interchangeably used the terms radiation necrosis and pseudoprogression[33,34], this should be avoided as there are differences in the clinical and radiological course of the two entities[28] and the histopathological and molecular phenotype differ[35]. The CNN study and the FET study included a mixture of cases of pseudoprogression and radiation necrosis.

Over time: a longitudinal imaging series can be used to analyse pseudoprogression

Dictionary learning has been employed to differentiate progression from pseudoprogression by performing implicit feature engineering without the need for tumour segmentation. In one glioblastoma study, features were estimated by using spatio-temporal discriminative dictionary learning of longitudinal diffusor tensor imaging (DTI) images to determine the sparse coefficients that were not shared between those with progression and pseudoprogression[36]. Then, after applying a score to each coefficient, a feature set was selected by sequentially adding the highest scoring coefficients using 10-fold cross validation and classifying the cases using an SVM. The best performance gave an accuracy of 0.87 and an AUC of 0.92. Again, it was unclear whether second line agents had been used, and there was no test dataset to validate the model. However, they were able to demonstrate some interpretability in that those with progression represented higher fractional anisotropy as might be expected due to the orientation of overproduced extracellular matrix in glioblastoma. Translation may be challenging because multiple concatenated DTI time points were required for the optimal classifier which might be logistically difficult to obtain in routine practice and again it is noteworthy that simple clinical features were not included.

Combinations: multiple imaging types can be combined as a means to analyse pseudoprogression

Traditional explicit feature engineering was used to differentiate progression from pseudoprogression within 3 months following chemoradiation of glioblastoma using simple and first order 3D shape features[37]. Post-contrast $T_1$-weighted and FLAIR images were combined, applying SVM and 4-fold cross validation. Sixty features were reduced to 5 and gave an accuracy of 0.9 in both a training dataset of 59 patients and a test dataset of 41 patients, which achieved 100% recall. Correlation coefficients comparing the most discriminant features at the two sites were high. The $T_2$-weighted hyperintensity phenotype of those patients with progression compared to those with pseudoprogression was round rather than elliptic; the post-contrast $T_1$-weighted phenotype was round and compact. As with the longitudinal DTI study, clinical data was not included in the analysis, and the results were not compared with simpler models, but the use of routine post-contrast $T_1$-weighted and $T_2$-weighted images increases the chance of translation.

Old and new: long-established ML methods have been used with advanced imaging to analyse pseudoprogression

As an alternative to SVM, a generalized linear model was applied to first order, second order and wavelet-transformed imaging features to differentiate progression from pseudoprogression in glioblastoma[38]. Post-contrast $T_1$-weighted, FLAIR, DSC and apparent diffusion coefficient (ADC) images were obtained within 3 months following chemoradiation from a training dataset of 61 patients. Feature selection by LASSO using 10-fold cross validation reduced the features from 6472 to 12. Classification using a generalized linear model showed that a multiparametric model of predominantly second order features (texture) gave an AUC of 0.90. Although relevant clinical and molecular data were collected, they were not included in any model despite MGMT promoter methylation status being shown to be significantly different in those with progression and pseudoprogression. The work was validated in a test dataset of 34 patients from a second hospital, although with a reduced AUC and accuracy, with some evidence of overfitting in the DSC component. This is likely to be associated with variation in how DSC is performed between centres[39], and is one reason why multiparametric techniques are challenging to translate.

Other long-established regression analyses within the definition of ML include multivariate logistic regression which has been employed in studies aiming to differentiate progression from pseudoprogression in glioblastoma[40-43]. A multivariate logistic regression model (LRM) employing LOOCV was applied in a study using DTI and DSC metrics to differentiate tissue containing pseudoprogression from tissue containing progression within 6 months following chemoradiation[41]. Using maximum relative cerebral blood volume (i.e. normalised to contralateral white matter; rCBV) and fractional anisotropy features obtained from the segmented enlarging contrast-enhancing lesions of 41 patients, the LRM gave an AUC of 0.81, recall of 0.79 and accuracy of 0.63.

LRM with LOOCV was also applied to 33 patients using dynamic contrast-enhanced imaging (DCE) metrics acquired from the enlarging contrast-enhancing lesion within 2 months after chemoradiation[42]. Unlike the other neuro-oncology monitoring studies in this review, this study was an entirely prospective study. Key clinical predictors were analysed and shown not to be discriminative. There was good inter-observer reliability. Using mean $K^{trans}$ (the volume transfer constant is a measure of capillary permeability obtained using DCE which reflects the efflux rate of gadolinium contrast from blood plasma into the tissue extravascular extracellular space), the LRM gave an accuracy of 0.76 and recall of 0.59.

In a further study of 35 patients, LRM was applied to subtracted ADC and DSC histograms of contrast-enhancing lesions obtained at baseline (around the time of chemoradiation) and at the point of enlargement within 6 months after chemoradiation[40]. Using the mode rCBV, LRM gave an AUC of 0.88, recall of 0.82 and accuracy of 0.94.

In summary, long-established ML methods can be used with advanced imaging techniques such as DSC or DCE to differentiate progression and pseudoprogression. A strength of the LRM studies is that the results are interpretable as they relate to the increased perfusion (CBV) and permeability ($K^{trans}$) occurring as a result of increased angiogenesis, the orientation of overproduced extracellular matrix (fractional anisotropy) and increased cellularity (ADC) known to be present in the enhancing rim of a glioblastoma. However, unlike in the generalized linear model approach, there were no test datasets employed in these single centre LRM studies.

Clustered combinations: unsupervised analyses can also be applied to multiple imaging types to analyse either pseudoprogression or the broader group of treatment-related effects

An unsupervised volume-weighted, voxel-based, multiparametric clustering method was used to differentiate progression from pseudoprogression within 3 months following chemoradiation[44] as well as recurrence from radiation necrosis in enlarging contrast-enhancing lesions seen after 3 months[45]. Pseudoprogression can occur up to 6 months[46], so the classifier in the second study is not examining radiation necrosis alone but two distinct entities combined[35] (or "treatment-related effects"). In the first study, metrics from ADC, DSC and DCE underwent *k*-means clustering in a training dataset of 108 patients and a test dataset of 54 patients. AUC in the test dataset was > 0.94 and accuracy and recall was > 0.87 for each of two readers with reliability intra-class correlation coefficient of 0.89. In the second study, the same metrics were included although a necrosis cluster was added to the finalized clusters analysed in the previous study. Boot strapping with LOOCV and 5-fold cross validation were used for evaluation. AUC in the training dataset of 75 patients was > 0.94 and recall was > 0.95 for each of two readers but there was no separate test dataset. As with many neuro-oncology monitoring biomarker studies, including the three studies using LRM above, it was unclear whether second line agents had been used which may confound the results. The results are impressive, particularly in the test dataset in the first study, however as with the LRM studies, multiparametric techniques are challenging to translate particularly with the known variation in advanced imaging techniques, including DCE, between centres[39].

Combinations and radiation necrosis: multiple imaging types can be combined as a means to analyse radiation necrosis

A feasibility study to differentiate radiation necrosis and progression in enlarging contrast-enhancing lesions seen after 9 months of chemoradiation was performed for both glioblastoma and brain metastases using FLAIR, $T_2$-weighted and contrast-enhanced $T_1$-weighted images[47]. There were 22 patients in a training dataset and 11 in a test dataset for glioblastoma patients and 21 in a training dataset and 4 in a test dataset for patients with brain metastases. Feature selection was performed with a feed-forward minimum redundancy and maximum relevance algorithm to reduce 119 features, including first and second order features as well as Laws and Laplacian pyramid features, to 5. Classification was performed by SVM recursive feature elimination with 3-fold cross validation. In the training datasets AUC was 0.79 for both tumour types using FLAIR images alone. In the test datasets, accuracy was 0.91 and 0.5 for glioblastoma and metastasis sub studies respectively, although all 3 MRI sequences were not available for all cases which makes interpretation challenging. The authors postulate that the features extracted in the study may relate to patterns similar to what is sometimes observed qualitatively in radiation necrosis, namely that the extracted Laws features relate to a soap-bubble appearance and that Laplacian pyramid features relate to an enhancing feathery rim. Furthermore, the Haralick features (second order texture features that are functions of the elements of the gray-level co-occurrence matrix and represent a specific relation between neighboring voxels) may relate to hypointensities and hyperintensities seen on all 3 MRI sequences due to microhaemorrhages in tumours. Because routine structural images were used, the chance of translation to the clinic is increased. Clinical data was not included in the analysis or models.

Voxel-based approaches can be used in the analysis of treatment-related effects

Proof-of-concept voxel-based approaches using ML to differentiate radiation necrosis and progression were developed in 2011 using DSC and ADC data[48]. In a recent study, whose aim was to differentiate progression from treatment-related effects (both pseudoprogression and radiation necrosis) in high grade glioma, a linear kernel SVM classifier was trained using DCE metrics (including $K^{trans}$) of 10 voxels within the enlarging contrast-enhancing lesion taken from 25 images from 20 patients[49]. Two-fold cross validation gave a recall of > 0.97. The model was applied to all voxels from a larger dataset of 44 images from the same 20 patients and shown to be interpretable and meaningful, including when there was a locally different treatment response in different lesions in the same patient. However, translation of the model may be challenging because it was trained on a small number of patients incorporating mixed grade, mixed treatment-effect (pseudoprogression and radiation necrosis), and mixed time points of the enlarging contrast-enhancing lesion (i.e. images not only from the first time point that an enlarging lesion is seen). There is also the potential for overfitting because images from several time points were used from the same patient to train the model.

Analysis of complete response

One study aimed to differentiate a complete response from progression a month before routine imaging assessment[3,4] would detect this using data from two immunotherapy studies[50]. Immunotherapy was added to the standard of care in one study and as a second line therapy in another. First and second-order features were extracted from FLAIR, $T_2$-weighted, contrast-enhanced $T_1$-weighted images, and other metrics were obtained from DTI and DSC images. Feature selection was performed using several algorithms including minimum redundancy maximum relevance and random forest to reduce 1248 features to 10 or less features. Classification was also performed by a range of algorithms and these included SVMs, random forest, linear discriminant analysis and stochastic gradient boosting. LOOCV, which consisted of leaving one patient out as opposed to one image out as multiple images were used for each patient, was performed during feature selection and classification. The highest balanced accuracy came from features derived from contrast-enhanced $T_1$-weighted and DSC images using a radial basis function SVM or boosting classifiers. However, no test dataset was used, and the methodology has significant weaknesses, in that it does not cater for a range of clinically-likely outcomes, such as stable disease.

## Prognostic Biomarkers

Prognostic biomarkers identify the likelihood of a clinical event, recurrence, or progression based on the natural history of the disease[1]. They are generally associated with specific outcomes such as overall survival or progression-free survival. The potential for confounding in prognostic biomarker and monitoring biomarker studies overlaps. Both may be influenced by second line treatments and a range of clinical variables. Most studies leveraging ML (Table 2) are also performed in a single centre and are retrospective.

Diagnostic biomarkers (described in detail in the other Special Issue publication dedicated to the application of machine learning in glioma imaging) may predict molecular information within a tumour from the imaging. Examples include MGMT promoter methylation status, 1p/19q chromosome arm co-deletion status and isocitrate dehydrogenase (IDH) mutation status. It is noteworthy that because some molecular markers are prognostic biomarkers in the same way that histopathological grade is a prognostic biomarker, diagnostic biomarkers may be prognostic biomarkers using the molecular marker or grade as a common biomarker. Another similarity of diagnostic and prognostic biomarker studies is that they both typically extract features from pre-operative MRIs and they often share methodology.

Given the overlap in principles described here and in the adjoining publication, we describe just two instructive studies as examples. In one study, an ML algorithm aimed to determine overall survival using imaging features from pre-operative routine MRI in patients with glioblastoma[51]. Pre- and post-contrast $T_1$-weighted, FLAIR, DSC and DTI images were obtained from a training dataset of 105 patients. Enhancing tumour tissue, non-enhancing tumour tissue, and oedematous tissue regions were segmented to produce imaging descriptors including location and first order statistics features and added to limited demographic features. Sixty features with the best survival prediction following 10-fold cross validation were selected from > 150 extracted features. Two SVMs were used to classify patients as survivors or not at 6 and 18 months respectively and a combined prediction index calculated. Ten-fold cross validation was used and gave an accuracy of 0.77 for predicting short, medium and long-term survivors. A prospective test dataset of 29 patients gave an accuracy of 0.79. Again, simple data such as performance status, which is known to be an important co-variate in multivariate analyses of glioma survival, were not included. To make the findings interpretable and meaningful, histograms were produced in order to understand the predictive features. Older patients, large tumour size, increased tumour diffusivity (potentially representing necrosis), larger proportions of $T_2$ hypointensity within a region, and highest perfusion peak heights, were all predictive of short survival. Although the findings have a plausible biological basis, translation is limited as this was performed in a single centre. It is also noteworthy that the process of predicting survival at set time points (6 and 18 months) is generally less useful than producing estimates over time (as survival curves allow).

An ML algorithm was used to determine overall survival of patients with high grade glioma using data from the brain tumor segmentation challenge (BRaTS)[52]. Pre- and post-contrast $T_1$-weighted, $T_2$-weighted and FLAIR images were obtained from a training dataset of 163 patients. Segmented regions including enhancing tumour tissue, non-enhancing tumour tissue, and oedematous tissue regions were manually segmented. Different sets of features were selected for classification. These included simple features such as location; histograms; discrete wavelet transform first and second order statistics; and a CNN which produced over 4000 deep features. The CNN was built using transfer learning based on AlexNet (a convolutional neural network that is trained on more than a million images from the ImageNet database[53]), and so benefits from the work already undertaken as part of the construction of an open source 'off-the-shelf' algorithm. Patients were classified as survivors or not at 10 and 15 months respectively. SVMs, k-nearest neighbours, linear discriminant analysis, tree, ensemble, and logistic regression were all independently applied to each set of features. A combination of CNN deep features and a linear discriminant classifier with 5-fold cross validation gave the best predictive result, although the reduction in accuracy between the training and test dataset (0.99 to 0.55) provides clear evidence of overfitting.

**Predictive Biomarkers**

Predictive biomarkers identify individuals likely to experience a favourable or unfavourable effect from a specific intervention or exposure[1]. Therefore, a predictive biomarker requires an interaction between treatment and the biomarker. Biological subsets (such as MGMT promoter methylation status, 1p/19q chromosome arm co-deletion status and IDH mutation status) may correlate with a favourable or unfavourable effect and in these cases there is an overlap with diagnostic and prognostic biomarkers[54]. There are few truly predictive biomarkers in neuro-oncology, molecular or otherwise. One study has applied unsupervised and supervised ML techniques to genomic information to predict whether pseudoprogression or true progression will occur after treatment[55]. Analytical and clinical validation in this radiogenomic study strongly suggested that interferon regulatory factor (IRF9) and X-ray repair cross-complementing gene (XRCC1), which were involved in cancer suppression and prevention respectively, are predictive biomarkers.

**Conclusion**

ML applications to imaging in neuro-oncology are at an early stage of development and applied techniques are not ready to be incorporated into the clinic. Many ML studies would benefit from improvements to their methodology. Examples include the use of larger datasets, the use of external validation datasets and comparison of the novel approach to simpler standard approaches. Initiatives and consensus statements have provided recommended frameworks[17,57,58] for standardizing imaging biomarker discovery, analytical validation and clinical validation which can help to improve the application of ML to neuro-oncology.

Studies taking advantage of enhanced computational processing power to build neuro-oncology monitoring biomarker models, for example using CNNs, have yet to show benefit compared to ML techniques using explicit feature engineering and less computationally expensive classifiers, for example using multivariate logistic regression. It is also notable that studies applying ML to build neuro-oncology monitoring biomarker models have yet to show overall advantage over those using traditional statistical methods[59,60]. However, regardless of method, increased computational power and advances in database curation will facilitate integration of imaging data with demographic, clinical and molecular marker data.

MRI is typically used throughout the neuro-oncology patient pathway however a major stumbling block of MRI is its flexibility. The same flexibility that makes MRI so powerful and versatile, also makes it hard to harmonise images from different centres. Afterall, MRI physics is complex and it is challenging (if not impossible) to fully harmonise parameters from different sequences, manufacturers and coils. These problems can be mitigated to some extent by manipulating the training dataset, such as through data augmentation, thereby allowing more generalisable ML models to be applied to MRI. Other approaches can describe the disharmony through modelling prediction uncertainty including the generation of algorithms that would "know when they don't know" what to predict.

Development and validation of ML models applied to neuro-oncology require large, well-annotated datasets, and therefore multidisciplinary and multi-centre collaborations are necessary. Radiologists are critical in determining key clinical questions and shaping research studies that are clinically valid. When these models are ready for the clinic as a routine clinical tool, as with the application of any medical device or the introduction of any therapeutic agent, there needs to be judicious patient and imaging selection reflecting the cohort used for validation of the model.

Alongside the drive towards clinical utility, the related issue of interpretability is likely to be important. As well as increasing user confidence, interpretability might help to generate new biological research hypotheses derived from image feature discovery.

## Acknowledgments

Funding: BLINDED

Table 1. Recent studies applying machine learning to the development of neuro-oncology monitoring biomarkers

| Author(s) | Prediction | Dataset | Method | Results |
|---|---|---|---|---|
| Cha J et al., 2014[40] | True progression | 35 CBV & ADC | Retrospective. Multivariate logistic regression, longitudinal subtraction of ADC & CBV histograms | Mode of rCBV AUC - 0.877 |
| Park JE et al., 2015[44] | Early true progression | 162 (training = 108 & testing = 54) DWI, DSC, DCE | Retrospective. Volume-weighted, MP clustering | Sensitivity - 87% Specificity - 87.1% AUC - 0.96 |
| Yun TJ et al., 2015[42] | True progression | 33 DCE | Prospective. Multivariate logistic regression, $K_{trans}$, $v_e$, $v_p$ | $K_{trans}$ Accuracy - 76% Sensitivity - 59% Specificity - 94% |
| Artzi M et al., 2016[49] | Pseudo progression | 20 longitudinal patients DCE & MRS (training = 25/44 DCE & MRS studies; testing = 19/44 studies) | Prospective. Voxel-wise SVM with $K_{trans}$, $v_e$, $K_{ep}$, $v_p$. | Sensitivity - 98% Specificity - 97% |
| Tiwari P et al., 2016[47] | Radiation necrosis | 58 (training = 43 & testing = 15) MRI | Retrospective. 119 features, mRmR feature selection, SVM. Sequence independent. | AUC - 0.79 AUC (primary) - 0.77 AUC (metastatic) - 0.72 |
| Qian X et al., 2016[36] | True progression | 35 longitudinal DTI | Retrospective. Spatio-temporal dictionary learning & SVM classification | Accuracy - 86.7% AUC - 0.92 |
| Ion-Margineanu A et al., 2016[50] | True progression | 29 $T_1$, $T_1$ C, DKI, DSC | Prospective. Compared 7 classifiers over various global and local features. | $T_1$ C Max BAR (balanced accuracy rate) value - 0.96 for AdaBoost |
| Yoon RG et al., 2017[45] | True progression | 75 MRI, DWI, DSC, DCE | Retrospective. Unsupervised. MP clustering of ADC, rCBV, IAUC | Sensitivity - 96.4% Specificity - 81.8% AUC - 0.95 |
| Booth TC et al., 2017[12] | True progression | 50 feature estimation. 24 (training = 17 & testing = 7) $T_2$ | Prospective testing set. SVM using Minkowski functionals | Accuracy - 88% AUC - 0.9 |
| Kebir S et al., 2017[30] | True progression | 14 18F-FET-PET | Retrospective. Unsupervised. Consensus | Sensitivity - 90% Specificity - 75% |

| Author(s) | | Dataset | Method | Results |
|---|---|---|---|---|
| | | | clustering, 19 conventional and textural features | NPV - 75% |
| Nam JG et al., 2017[43] | True progression | 37 DCE | Retrospective. Multivariate logistic regression using pharmacokinetic parameters | $K_{ep}$ Accuracy - 70.3% AUC - 0.75 Sensitivity - 71.4% Specificity - 90.0% |
| Jang B-S et al., 2018[31] | Pseudo progression | 78 (training = 59 & testing = 19) $T_1$ C MRI, Age, Gender, MGMT status, IDH mutation, radiotherapy dose & fractions, follow up interval | Retrospective. 9 $T_1$ C axial slices centred on lesion, CNN | AUC - 0.83 AUPRC - 0.87 F1 score - 0.74 |
| Ismail M et al., 2018[37] | True progression | 105 (training = 59 & testing = 46) MRI | Retrospective. SVM using global & local features of lesion & peritumour habitat | Accuracy - 90.2% Sensitivity - 100% Specificity - 94.7% |
| Kim JY et al., 2018[38] | Early true progression | 95 (training = 61 & testing = 34) $T_1$ C, FLAIR, DWI, DSC | Retrospective. Generalised linear model, LASSO feature selection on multiparametric first- & second-order statistics. | AUC - 0.85 Sensitivity - 71.4% Specificity - 90.0% |

18F-FET-PET = [18F]-fluoroethyl-L-tyrosine positron emission tomography. NPV = negative predictive value. $T_1$ C = post contrast $T_1$-weighted. MGMT = $O^6$-methylguanine-DNA methyltransferase. IDH = isocitrate dehydrogenase. CNN = convolutional neural network. AUC = area under the receiver operator curve. AUPRC = area under the precision-recall curve. DCE = dynamic contrast-enhanced imaging. MRS = 1H-magnetic resonance spectroscopy. SVM = support vector machine. mRmR = minimum redundancy and maximum relevance. CBV = cerebral blood volume (rCBV =relative CBV). ADC = apparent diffusion coefficient. IAUC = initial area under the curve. MP =multi parametric. DWI = diffusion weighted imaging. DSC = dynamic susceptibility-weighted. LASSO = least absolute shrinkage and selection operator. DTI = diffusor tensor imaging. DKI = diffusor kurtosis imaging.

Table 2. Recent studies applying machine learning to the development of neuro-oncology prognostic biomarkers

| Author(s) | Dataset | Method | Results |
|---|---|---|---|
| Choi YS et al., 2015[61] | 61 preoperative DCE | Retrospective. Multivariate Cox regression using MRI, pharmacokinetic, & clinical parameters | C-index - 0.82 |
| Kickingereder P et al., 2016[62] | 119 (training = 79 & testing = 40) $T_1$, $T_1$ C, FLAIR, DWI, DSC | Retrospective. Supervised Principal Component Analysis with Cox regression analysis | C-index - 0.70 |
| Chang K et al., 2016[63] | 126 (training = 84 & testing = 42) patients $T_1$, $T_2$, FLAIR, $T_1$ C, DWI | Retrospective. Random forest on radiomic features (including Laws, Haralick) | Accuracy - 76% |
| Liu L et al, 2016[64] | 147 rs-fMRI and DTI | Retrospective. SVM using clinical features & network features of structural & functional network | Accuracy - 75% |
| Nie D et al, 2016[65] | 69 $T_1$ C, rs-fMRI, DTI | Prospective. SVM using supervised CNN-derived features | Accuracy - 89.9% Sensitivity - 96.9% Specificity - 83.8% PPR - 84.9% NPR - 93.9% |
| Macyszyn L et al., 2016[51] | 134 (training = 105 & testing = 29) $T_1$, $T_1$ C, $T_2$, FLAIR, DTI, DSC | Prospective. SVM for OS < 6 months & SVM for OS < 18 months | Accuracy (< 6 months) - 82.76% |

| | | | Accuracy (< 18 months) - 83.33% Accuracy (combined) - 79% |
|---|---|---|---|
| Zhou M et al., 2017[66] | 32 TCGA $T_1$ C, FLAIR, $T_2$ & 22 $T_1$ C, FLAIR, $T_2$ | Retrospective. Group Difference Features to quantify habitat variation. Supervised forward feature ranking with SVM. | Accuracy - 87.5%, 86.4% |
| Dehkordi ANV et al., 2017[67] | 33 pre-treatment DCE | Retrospective. Adaptive Neural Network with Fuzzy Inference System using $K_{trans}$, $K_{ep}$ and $v_e$ | Accuracy - 84.8% |
| Lao J et al., 2017[68] | 112 (training = 75 & testing = 37) pre-treatment $T_1$, $T_1$ C, $T_2$, FLAIR | Retrospective. Multivariate Cox regression analysis using radiomic features as well as 'deep features' from pre-trained CNN. | C-index - 0.71 |
| Liu Y et al., 2017[69] | 133 $T_1$ C | Retrospective. Recursive Feature Selection with SVM | Accuracy - 78.2% AUC - 0.81 Sensitivity - 79.1% Specificity - 77.3% |
| Li Q et al., 2017[70] | 92 (training = 60, testing = 32) $T_1$, $T_1$ C, $T_2$, FLAIR. TCGA data used. | Retrospective. Random forest for segmentation into 5 classes. Multivariate LASSO-Cox regression model. | C-index - 0.71 |
| Chato L & Latifi S, 2017[52] | 163 $T_1$, $T_1$ C, $T_2$, FLAIR. Short-, mid-, long-term survivors | Retrospective. SVM, KNN, linear discriminant, tree, ensemble & logistic regression applied to volumetric, statistical & intensity texture, histograms & deep features. | Accuracy - 91% Linear discriminant using deep features |
| Ingrisch M et al., 2017[71] | 66 $T_1$ C | Retrospective. Random survival forests using 208 global & local features from segmented tumour | C-index - 0.67 |
| Li Z-C et al., 2017[72] | 92 (training = 60 & testing = 32) $T_1$, $T_1$ C, $T_2$, FLAIR. TCGA data used. | Retrospective. LASSO Cox regression to define radiomics signature | C-index - 0.71 |
| Bharath K et al., 2017[73] | 63 TCGA preoperative - $T_1$ C, FLAIR | Retrospective. LASSO Cox regression using Age, KPS, DDIT3 & 11 Principal Component shape coefficients | C-index - 0.86 |
| Z Shboul et al., 2017[74] | 163 $T_1$, $T_1$ C, $T_2$, FLAIR | Retrospective. Recursive feature selection & Random Forest regression. | Accuracy – 63% |
| Peeken JC et al., 2018[75] | 189 $T_1$, $T_1$ C, $T_2$, FLAIR & clinical data. | Retrospective. Multivariate Cox regression using VASARI features and clinical data | C-index - 0.69 |
| P Kickingereder et al., 2018[76] | 181 (training = 120 & testing = 61) pretreatment MRI | Retrospective. Penalized Cox model for radiomic signature construction. | C-index – 0.77 |
| A Chaddad et al., 2018[77] | 40 (training = 20 & testing = 20) preoperative MRI, $T_1$ & FLAIR. | Retrospective. Random Forest on multi-scale texture features | AUC – 74.4% |
| S Bae et al., 2018[78] | 217 (training = 163 & testing = 54) pre- | Retrospective. Variable hunting algorithm for selection & Random Forest classifier | iAUC - 0.65 |

| | operative MRI, $T_1$ C, $T_2$,, FLAIR, DWI | | |
|---|---|---|---|

TCGA = The Cancer Genome Atlas . $T_1$ C = post contrast $T_1$-weighted. SVM = support vector machine. DCE = dynamic contrast-enhanced imaging. CNN = convolutional neural network. KNN = k-nearest neighbours/ rs-fMRI = resting state functional MRI. KPS = Karnofsky performance status. DDIT3 = DNA Damage Inducible Transcript 3. DTI = diffusor tensor imaging. DSC = dynamic susceptibility-weighted. OS = overall survival

**Figure 1.** The phases of radiomics are shown using explicit feature engineering. Some pre-processing steps are shown here: manual segmentation of hyperintense voxels associated with a glioblastoma in a $T_2$-weighted image is performed. A mask is extracted which undergoes quantization. Some feature estimation steps are shown here: in this example, the pixels are made into three features which are topological descriptors of image heterogeneity[12] (area is the number of white pixels = 1; perimeter around a white pixel = 4; genus is the number of rings subtracted from number of holes = 0). Note that deep learning uses implicit feature engineering and some of the feature estimation steps may not be required.

**Figure 2.** A longitudinal series of $T_1$-weighted images after gadolinium administration. On the left is an image demonstrating a glioblastoma 1 month after surgery before chemoradiotherapy. In the middle is an image demonstrating the appearances 2 months after radiotherapy and concomitant chemotherapy. On the right is an image demonstrating the appearances 4 months after radiotherapy and concomitant chemotherapy. There was no new treatment between 2 and 4 months therefore this shows pseudoprogression occurred at 2 months.